# Soft modes and NTE in Zn(CN)$_2$ from Raman spectroscopy and first principles calculations

T. R. Ravindran[*], A. K. Arora, Sharat Chandra, M. C. Valsakumar and N. V. Chandra Shekar

Materials Science Division, Indira Gandhi Centre for Atomic Research, Kalpakkam 603 102, India

We have studied Zn(CN)$_2$ at high pressure using Raman spectroscopy, and report Gruneisen parameters of the soft phonons. The phonon frequencies and eigen vectors obtained from ab-initio calculations are used for the assignment of the observed phonon spectra. Out of the eleven zone-centre optical modes, six modes exhibit negative Gruneisen parameter. The calculations suggest that the soft phonons correspond to the librational and translational modes of C≡N rigid unit, with librational modes contributing more to thermal expansion. A rapid disordering of the lattice is found above 1.6 GPa from X-ray diffraction.

[*]Corresponding author: Email: trr@igcar.gov.in

Interest in materials that exhibit negative thermal expansion (NTE) was renewed after the report [1] of high and isotropic NTE in $Zr(WO_4)_2$ over a wide temperature range, leading to extensive work and several reviews on the subject [2-5]. The structure of $Zr(WO_4)_2$ and several other NTE materials consist of corner sharing tetrahedral and octahedral units. From Raman spectroscopic investigations on $Zr(WO_4)_2$ as a function of pressure and temperature, the phonons responsible for NTE have been identified, and it has been shown that in addition to the librational (rigid-unit) mode at 5 meV, several other phonons of much higher energy also contribute significantly to the NTE in this material [6-9]. Based on structural analysis transverse displacements of the shared oxygen atoms and consequent rotation of polyhedra [1] was suggested as the cause of NTE in $Zr(WO_4)_2$. In the context of corner linked structures, $Zn(CN)_2$ is remarkable, as it has C≡N as the linking species between tetrahedral units instead of a single atom and exhibits twice as much coefficient of NTE ($-17 \times 10^{-6}$ $K^{-1}$) [10] as that of $Zr(WO_4)_2$. The structure of $Zn(CN)_2$ consists of three-dimensional, inter-penetrating, tetrahedral frameworks of Zn-CN-Zn chains [11]. Two different cubic structures, $P\bar{4}3m$ (CN-ordered) [11] and $Pn\bar{3}m$ (CN-disordered) [10] have been reported to fit well to the diffraction patterns. In the 'ordered structure' the CN ions lying along the body diagonal are orientationally ordered such that they form $ZnC_4$ and $ZnN_4$ coordination tetrahedra around alternate cations. On the other hand, in the 'disordered structure', C and N atoms are randomly flipped so as to occupy the sites with equal probability. It was shown recently from factor group analysis in conjunction with Raman and IR spectroscopic measurements that the structure is indeed disordered [12].

From a topological model treating $ZnN_4/ZnC_4$ as rigid units the structure was argued to support a large number of low frequency rigid unit phonon modes ($\omega_{ph}$< 2 THz, ≈70 cm$^{-1}$) that contribute to NTE [13]. On the other hand, it was shown recently by spectroscopic measurements [12] that the lowest energy optical mode in $Zn(CN)_2$ is an IR mode at 178 cm$^{-1}$. It has been shown by atomic pair distribution function analysis of X-ray diffraction data and suitable modelling of the displacements of C/N away from the body-diagonal that this displacement increases as a function of temperature [14]. This is in any case expected from the increased amplitude of atomic vibrations as temperature is increased. However, there is no report of the role of different phonons to thermal expansion. Since phonon modes and their Gruneisen parameters are directly responsible for thermal expansion in a material, it becomes vitally important to study them.

Here we report the first study of phonons in $Zn(CN)_2$ at high pressure using Raman spectroscopy and ab-initio calculations. High pressure X-ray diffraction measurements are also carried out for obtaining the bulk modulus to calculate Gruneisen parameters. From high pressure Raman measurements soft phonons are identified. In addition, first-principles ab-initio density functional calculations are performed at different volumes and phonon dispersion curves obtained using frozen phonon approximation with SIESTA code [15]. The phonon eigen vectors are used for the assignment of phonon modes. The thermal expansion coefficient is calculated from Gruneisen parameters of all the phonons and compared with the reported value.

$Zn(CN)_2$ (>99.5%) was obtained from Alfa Aesar. X-ray diffraction pattern of this powder sample showed no observable impurity phases. A small piece of sample of lateral dimensions ~100 μm was loaded into a gasketed, Mao-Bell type diamond anvil cell. Raman spectra were recorded at different pressures in the backscattering geometry using the 488-nm line of an argon ion laser. Methanol-ethanol (4:1) mixture was used as pressure transmitting medium. Ruby fluorescence was used to measure pressure. Scattered light from the sample was analyzed by a SPEX double monochromator, and detected with a cooled photomultiplier tube operated in the photon counting mode. Scanning of the spectra and data acquisition were carried out using a PSoC (Programmable System on Chip) hardware controlled by LabVIEW® 7.1 program [16]. The spectral range covered was 10-2400 $cm^{-1}$ that also includes the C≡N stretch mode around 2220 $cm^{-1}$. High pressure X-ray diffraction (HPXRD) was carried out in an angle dispersive mode using Guinier diffractometer [17]. The incident Mo $K_{\alpha 1}$ radiation is obtained from a Rigaku 18 kW rotating anode x-ray generator.

Ab-initio calculations were carried out in the framework of the density functional theory using the Perdew–Burke–Ernzerhof generalized gradient approximation for exchange and correlation [18]. A 3×3×3 supercell of $Zn(CN)_2$ unit cell was used for determining the relaxed atomic configuration, and phonon frequencies calculated using the SIESTA code. The calculations were performed using a Monkhorst-Pack grid of 8×8×8 **k-**points with a shift of 0.5. The energy cut off was 350 Rydbergs and a double zeta plus polarization (DZP) basis set was used. Standard norm-conserving, fully relativistic Troullier-Martins TM2 pseudopotentials were used. The computations were performed in a 16-node linux cluster.

The 30 degrees of freedom arising from the 10 atoms in the cubic unit cell of $Zn(CN)_2$ result in 3 acoustic and 27 optical branches. Out of the three structural units viz., C≡N ion, $ZnC_4$ tetrahedron and $ZnN_4$ tetrahedron, C≡N is the most strongly bound unit and hence taken as a 'rigid molecular unit'. The 6 degrees of freedom corresponding to the linear molecular ion C≡N can be divided into 1-internal (stretching vibration), 3 rigid-translations and 2 rigid rotational degrees of freedom. The 'disordered' structure of zinc cyanide has the following irreducible representations of optical phonons [12]:

$$\Gamma^{Opt.} = A_{1g} + E_g + F_{1g} + 3F_{2g} + A_{2u} + E_u + 2F_{1u} + F_{2u}$$

Out of these, the $A_{1g}$, $E_g$ and $F_{2g}$ modes are Raman active and $F_{1u}$ mode is IR active. The remaining four modes are optically inactive.

Figure 1 shows the Raman spectra of $Zn(CN)_2$ at several pressures including ambient. There are three Raman modes clearly seen at 2221, 342, and 200 cm$^{-1}$. Out of these, the asymmetric peak about 342 cm$^{-1}$ is actually a doublet that can be resolved into 339 and 343 cm$^{-1}$ [12]. The linewidth of all three modes increase and their intensities reduce at high pressures. While the C≡N stretch mode about 2220 cm$^{-1}$ hardens as pressure is increased, the other two modes are seen to soften. The modes at 342 cm$^{-1}$ and 200 cm$^{-1}$ are too weak to follow above 1 GPa. Figure 2 depicts the phonon frequency ($\omega$) vs. pressure ($P$) for the three modes observed by Raman spectroscopy. Most measurements were carried out under hydrostatic conditions using (methanol + ethanol) as pressure transmitting medium. One set of measurements in which no medium was used (open circles in Fig. 2) resulted in a weaker $P$ dependence for the 2220 cm$^{-1}$ mode up to a pressure of 1.5 GPa and a negative coefficient above 2 GPa. The reason for this change of slope from positive to negative could be a structural transition occurring under non-hydrostatic pressure. However, high pressure X-ray diffraction measurements (discussed later) under hydrostatic pressure have not indicated any structural phase transition between 1.5 and 2 GPa or at any pressure up to 5.2 GPa, the highest pressure up to which measurements were made.

X-ray diffraction patterns were recorded at several pressures up to 5.2 GPa. Only three reflections, viz., (110), (211) and (321) could be observed. As pressure is increased, the intensity of all lines reduces drastically. At a pressure as low as 0.2 GPa the intensities of the peaks reduce by about 50%. Above 0.6 GPa the (321) line disappears and above 1.6 GPa, only the (110) line is present, which continues up to the highest pressure of 5.2 GPa, indicating possible disordering of C≡N has taken place above 1.6 GPa. Such a partial/sublattice amorphization has been reported earlier also in other compounds [19, 20]. Lattice parameters at several pressures were obtained from the three lines using a disordered cubic space group ($Pn\bar{3}m$) structure. The unit cell volume obtained as a function of pressure was fitted to Murnaghan equation of state and resulted in a bulk modulus $B_0$=25±11 GPa. The large error in $B_0$ is due to the scatter in the XRD data and also the small number of reflections that were used to calculate the lattice parameters. With this input of $B_0$ the mode Gruneisen parameters ($\gamma_i = B_0 \omega_i^{-1} \partial\omega_i/\partial P$) of the three Raman modes could be calculated (Table 1, last column). In the absence of $\gamma_i$ values of other vibrational modes, the thermal expansion coefficient has been calculated using simulation data as detailed in the next paragraph.

It is not straight forward to incorporate random disorder in ab-initio calculations. When such a disorder is introduced by randomly flipping half the C≡N species in the supercell, it is found that for this disordered structure of cubic $Zn(CN)_2$ - when the system is allowed to relax - the ground state energy does not converge to a stable configuration but evolves into a tetragonal structure (space group $P4_2nm$) with $c$-parameter ~0.5% larger than the $a$- and $b$-parameters. Upon further relaxation, the structure slowly becomes triclinic. Additionally, the inter-atomic forces are large and do not converge to small values. On the other hand, for the 'ordered structure' the forces converged to values less than $10^{-6}$ eV/Å due to geometrical considerations. Hence the ordered structure of $Zn(CN)_2$ is used for computational purposes. It should be pointed out that the values of the vibrational frequencies obtained from either of the space groups are not expected to be different from each other, since the same kind of atomic motions are involved in the vibrational modes. The number of zone-centre optical phonon modes is also the same in either space group. The total energy of the system was computed in the relaxed configuration for different volumes of the cell up to $V/V_0$=0.844. The energy vs. volume data was fitted to Murnaghan equation and the bulk modulus obtained is 88 GPa. A similar result (90 GPa) is obtained when WIEN2K is used to calculate the bulk modulus. Phonon dispersion curves at different volumes were calculated using the frozen-phonon method using the VIBRA module in the SIESTA package. Eigen frequencies for the various modes were obtained by diagonalizing the dynamical matrix. The phonon dispersion curves obtained at ambient volume from simulations are shown in Figure 3. Eigenvectors were viewed using the Visual Molecular Dynamics (VMD) package [21]. The highest compression corresponds to a pressure of 8.3 GPa. From the pressure dependence of the various zone centre optical phonons (inset in Fig. 2) the mode Gruneisen parameters were obtained [Table I]. Using Einstein's specific heat $C_i = R [x_i^2 \exp(x_i)]/[\exp(x_i)-1]^2$, where $x_i=\hbar\omega_i/k_BT$, for the various modes the total specific heat $C_V$ was obtained. Here $R$ is the universal gas constant. Thermal expansion coefficient $\alpha=(\gamma_{av}C_V)/(3V_mB_0)$, (where $\gamma_{av}=\frac{1}{2}\sum p_iC_i\gamma_i)/C_V$, $p_i$ are the degeneracies of the respective $\omega_i$ phonon branches at the Brillouin zone centre, $V_m$ is the molar volume and $B_0$ taken as 88 GPa) is calculated to be $-22\times10^{-6}$ K$^{-1}$, in good agreement with the reported value.

In view of the non-availability of polarized Raman measurements on oriented single crystals of Zn(CN)$_2$, the observed modes were assigned (Table I) based on eigen vectors of calculated phonons. The CN stretching mode at 2200 cm$^{-1}$ can be assigned to $A_{1g}$. In the internal mode region $A_{1g}$ and $F_{2g}$ modes arise due to correlation splitting [12] and are often degenerate. It is noteworthy that six out of the eleven optical modes exhibit negative Gruneisen parameters. Furthermore, all the modes of energy lower than 360 cm$^{-1}$ have negative $\gamma_i$. Figure 4 shows the displacement vectors of different atoms for the phonons that exhibit large negative $\gamma_i$. The 143 cm$^{-1}$ mode corresponds to translational motion of CN ions whereas the other three modes involve librations of CN ions about the axis joining Zn-Zn' atoms. The difference in the values of the calculated and the observed values of $\gamma_i$ could partly arise from the different values of $B_0$ used. However, this does not affect the calculation of $\alpha$, since $B_0$ gets cancelled in the definition of $\alpha$. Further, the total Gruneisen parameter ($\Sigma p_i \gamma_i$) for the C≡N librational modes is -57 whereas for the translational modes this value is -41.

As mentioned earlier, using a topological model the network structure of Zn(CN)$_2$ has been argued to have a large number of low-frequency rigid units modes of ZnC$_4$/ZnN$_4$ in analogy with Zr(WO$_4$)$_2$. On the other hand, in the present lattice dynamical calculations the lowest frequency mode turns out to be a CN-translational mode. This is because the topological model treated ZnC$_4$/ZnN$_4$ as rigid units, whereas actually only the strongly bound CN ions should be considered as rigid units. Recent atomic pair distribution function analysis shows that the displacements of C/N away from the line joining Zn…Zn' increases as a function of temperature. Though this appears physically reasonable, this displacement, when extrapolated to 0 K, remains as large as 0.42 Å (Fig.9 of Ref.[14]) suggesting inconsistency between the Rietweld refined structure and that obtained from PDF analysis. Furthermore, the reason for Zn…Zn' distance (which is directly related to the lattice parameters) estimated from PDF analysis being different from that obtained from XRD analysis [14] remains unclear. On the other hand, the present phonon calculations and Raman measurements at high pressure provide the first insight into the relative role of the different phonons in causing negative thermal expansion in $Zn(CN)_2$.

In conclusion, we have identified the optical phonons responsible for NTE in $Zn(CN)_2$ from high pressure Raman spectroscopic studies and from first principles density functional simulation studies at different volumes. Gruneisen parameters of all the vibrational modes were obtained from simulations. A large number of phonon modes in $Zn(CN)_2$ are soft, and all contribution to NTE arises from C≡N librational and translational modes. The value of thermal expansion coefficient α calculated from the Gruneisen parameters is in good agreement with experimental value. X-ray diffraction investigations suggest growth of disorder at high pressure.

Table I

| Modes (deg. of freedom) | Symmetry | Calc. freq. (cm$^{-1}$) | Obs. freq. (cm$^{-1}$) | Calc. $\gamma_i$ | Obs. $\gamma_i$ |
|---|---|---|---|---|---|
| Zn-trans. (3) | $F_{2g}$ | 388 | 343 (R) | 0.45 | - |
| CN-trans. (12) | $F_{1u}$ | 143 | 178 (IR) | -14.3 | - |
|  | $E_u$ | 255 | Inactive | -1.5 | - |
|  | $F_{2g}$ | 352 | 216 (R) | -0.13 | -0.50 (15) |
|  | $A_{2u}$ | 564 | Inactive | 1.1 | - |
|  | $F_{1u}$ | 596 | 461 (IR) | 1.4 | - |
| CN-libr. (8) | $F_{1g}$ | 288 | Inactive | -8.0 | - |
|  | $E_g$ | 357 | 339 (R) | -6.2 | -0.54 (2) |
|  | $F_{2u}$ | 326 | Inactive | -7 | - |
| CN-int. (4) | $F_{2g}$ | 2232 | 2218 (IR) | 1.5 | 0.14(1) |
|  | $A_{1g}$ | 2245 | 2221 (R) | 1.5 | - |

**Figure and table captions:**

Table 1. Calculated and observed phonon frequencies in $Zn(CN)_2$, their classification, mode assignments and Gruneisen parameters. Observed IR frequencies are from [12].

Figure 1. Raman spectra of $Zn(CN)_2$ at several pressures. Spectra are scaled and shifted for clarity. The modes at 342 and 200 cm$^{-1}$ could not be followed above 1 GPa due to weak intensities.

Figure 2. Mode frequency vs. Pressure for the observed Raman modes in $Zn(CN)_2$. Open symbols: results without pressure medium. The inset shows $\omega$ vs. *P* for all the eleven modes obtained from the phonon calculations. Though data were generated up to 8.3 GPa, the trend after the first three pressures (shown here) is non-linear, and hence not considered for obtaining $\gamma_i$.

Figure 3. Phonon dispersion curves obtained from First Principles density functional simulations on a 3×3×3 supercell of $Zn(CN)_2$. Note that the acoustic phonon branch interacts with the lowest energy optical phonon branch at 143 cm$^{-1}$. Both the branches change character due to the non-crossing rule.

Figure 4. Atomic displacements of vibrational modes corresponding to (a) 143 cm$^{-1}$, (b) 288 cm$^{-1}$, (c) 326 cm$^{-1}$ and (d) 357 cm$^{-1}$. The arrows fixed to the atoms are proportional to the amplitude of atomic motion. In the 326 cm$^{-1}$ mode neighbouring Zn atoms also move (opposite direction)

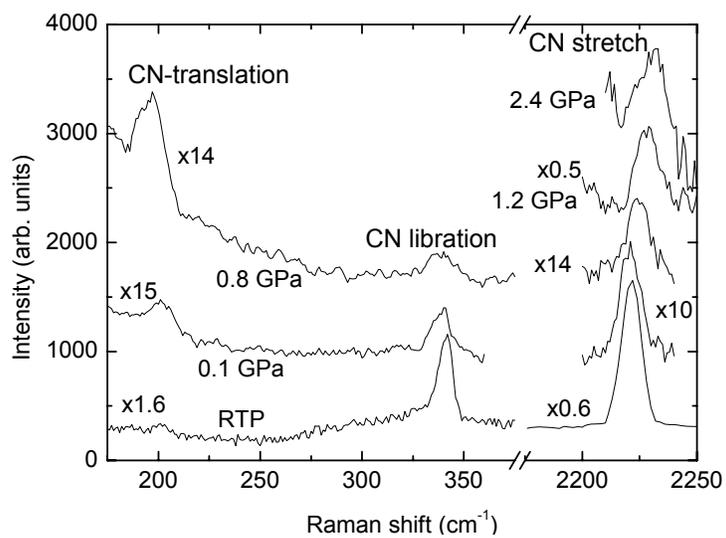

Figure 1. Ravindran et al

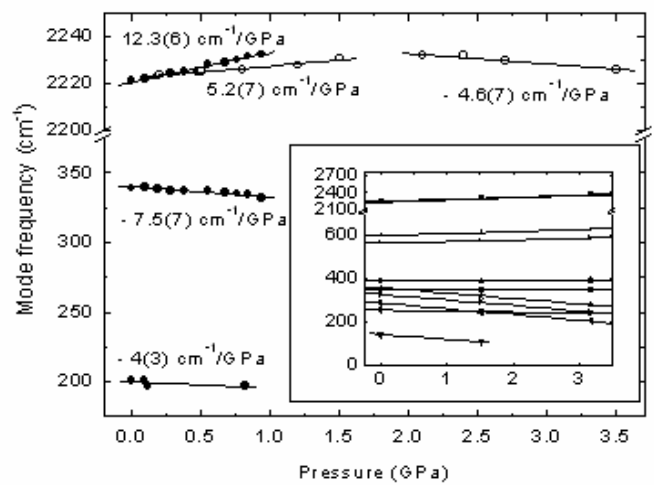

Figure 2. Ravindran et al.

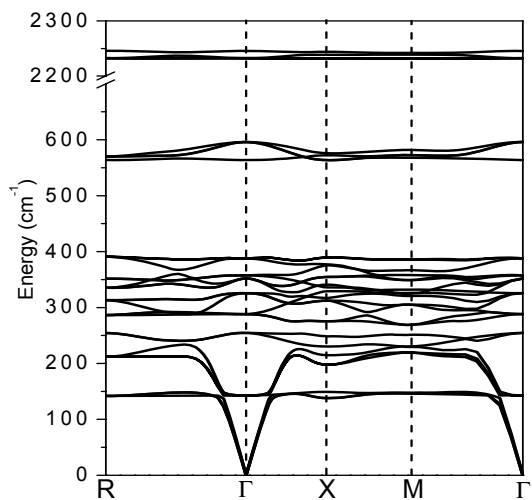

Figure 3. Ravindran et al.

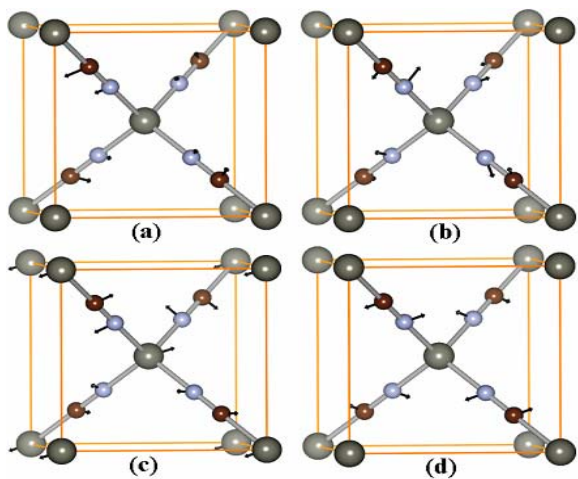

Figure 4. Ravindran et al.

15